\documentclass[12pt,a4paper]{article}
\usepackage{amsfonts,latexsym}
\usepackage{amsmath,amssymb}
\usepackage{graphicx,color}
\usepackage{ulem}

\renewcommand{\thefootnote}{\fnsymbol{footnote}}

\begin{document}

\vspace{12mm}

\begin{center}
{{{\Large {\bf Simplest  model of a scalarized black hole  in the Einstein-Klein-Gordon theory }}}}\\[10mm]

{Xiao Yan Chew$^1$\footnote{e-mail address: xiao.yan.chew@just.edu.cn} and Yun Soo Myung$^{2}$\footnote{e-mail address: ysmyung@inje.ac.kr}}\\[8mm]

{${}^1$School of Science, Jiangsu University of Science and Technology, Zhenjiang, 212100 China\\[0pt]}

{${}^2$Institute of Basic Sciences and Department  of Computer Simulation,  Inje University, Gimhae 50834, Korea\\[0pt]}

\end{center}

\vspace{2mm}

\begin{abstract}
We investigate scalarized black holes in the Einstein-minimally coupled scalar theory with a negative potential $V(\phi)=-\alpha^2\phi^6$.
The tachyonic instability is absent  from analyzing the linearized scalar equation, which could  not  allow for spontaneous scalarization.
However, we obtain the black hole solutions with scalar hair  by solving three full equations because this scalar potential violates the weak energy condition.
This shows clearly that a single branch of scalarized black holes can be obtained without introducing a non-minimal scalar coupling term.
We perform the stability  analysis  for scalarized black holes by adopting radial perturbations,
implying that  all  scalarized black holes belonging to a single branch are unstable.
\end{abstract}
\vspace{5mm}

\newpage
\renewcommand{\thefootnote}{\arabic{footnote}}
\setcounter{footnote}{0}

%%%% Introduction %%%%

\section{Introduction}

The no-hair theorem in general relativity (GR) prevents the existence of asymptotically flat black hole solutions with scalar hair
except the mass $M$, angular momentum $J$, and charge $Q$ of the black hole~\cite{Israel:1967wq,Carter:1971zc,Ruffini:1971bza}.
This is based on the Einstein-minimally coupled scalar theory~\cite{Herdeiro:2015waa}.

If one introduces a conformally coupled scalar (Einstein-conformally coupled scalar theory), it gave  us the BBMB black holes with a conformal scalar hair, indicating  an evasion of no-hair theorem~\cite{Bocharova:1970skc,Bekenstein:1974sf}. In this case, however,  the conformal scalar hair blows up at the horizon and  these black holes are unstable under linear perturbations.

Recently, the spontaneous scalarization implies that  infinite branches of scalarized (charged) black holes were obtained numerically  from the Einstein-Gauss-Bonnet-scalar theory~\cite{Antoniou:2017acq,Doneva:2017bvd,Silva:2017uqg,Blazquez-Salcedo:2018jnn} (Einstein-Maxwell-scalar theory~\cite{Herdeiro:2018wub})
through the non-minimal coupling function $f(\phi)$ with $f(0)=0,f'(0)=0,f''(0)\not=0$  to the Gauss-Bonnet term (Maxwell term). In these linearized theories ($\bar{\nabla}^2\delta\phi-m^2_{\rm eff}\delta\phi=0$), the tachyonic instability of $s(l=0)$-scalar
mode propagating around the GR black holes indicates the onset of spontaneous scalarization. This arose  from either the negative Gauss-Bonnet coupling term ($m^2_{\rm eff}=-48\lambda^2 M^2/r^6$) or the negative Maxwell coupling term ($m^2_{\rm eff}=-2\alpha Q^2/r^4$) where $\lambda^2$ and $\alpha$ are positive scalar coupling parameters.  These negative coupling terms induce actually potential wells near the horizon and as coupling parameters increase, leading to tachyonic instabilities.  Furthermore,
the linearized  static scalar equation has played an important role in obtaining  scalar clouds (bound states). Requiring an asymptotically vanishing scalar, the condition for obtaining a smooth scalar selects a discrete set of the bifurcation points for scalarized solutions: $M/\lambda=\{0.587,0.226,0.140,\cdots\}~(\alpha(q=0.7)=\{8.019,40.84,99.89,\cdots\})$ for the Gauss-Bonnet term~\cite{Myung:2018iyq} (as well as the Maxwell term~\cite{Myung:2018vug}).
 They have  admitted  scalar hairy black holes in the full theory. Actually, it has induced  infinite branches of  scalarized [charged]  black holes: $n=0(0<M/\lambda<0.587),1(0<M/\lambda<0.226),2(0<M/\lambda<0.140),\cdots[n=0(\alpha>8.019),n=1(\alpha>40.84),n=2(\alpha>99.89),\cdots]$.

However, it turned out that the $n=0$ branch of scalarized (charged) black holes are stable against radial perturbations, whereas all other branches of $n\not=0$ are unstable~\cite{Zou:2020zxq}.
This implies that the $n=0$ branch of scalarized (charged) black holes  could survive for  further implications of scalarized black holes.  In this direction, the dynamics of scalarized  black holes and binary mergers were addressed in the  Einstein-Gauss-Bonnet-scalar theory~\cite{Witek:2018dmd,Silva:2020omi,Kuan:2021lol,East:2021bqk,Blazquez-Salcedo:2020caw,Staykov:2021dcj,Blazquez-Salcedo:2022omw}. Furthermore, the photon spheres and observational appearance  of scalarized  black holes
are investigated in the Einstein-Maxwell-scalar theory~\cite{Konoplya:2019goy,Blazquez-Salcedo:2019nwd,Astefanesei:2019qsg,LuisBlazquez-Salcedo:2020rqp,Gan:2021pwu,Gan:2021xdl,Li:2024oyc,Sui:2023yay}.

On the other hand, the nonlinear mechanism was introduced to obtain a single branch of  scalarized black holes in Einstein-scalar-Gauss-Bonnet gravity, which is surely beyond the previous spontaneous scalarization.
Introducing  a different coupling function $f(\phi)$ satisfying $f(0)=0,f'(0)=0,f''(0)=0$,  one found from its linearized equation ($\bar{\nabla}^2\delta\phi=0$) that Schwarzschild black hole is linearly stable against scalar perturbation, whereas it is  unstable against nonlinear scalar perturbation when the amplitude of a perturbed scalar  is large enough~\cite{Blazquez-Salcedo:2022omw,Doneva:2021tvn,Pombo:2023lxg,Zhang:2023jei}.
This provides another mechanism to obtain a single branch of  scalarized black holes via nonlinear scalarization~\cite{Lai:2023gwe,Zhang:2024spn}.

In this work, we wish to introduce another model to find a single branch of scalarized black holes.  Its linearized theory is stable against  scalar perturbation, whereas its full theory  indicates a violation of the weak energy condition (WEC)
because the scalar potential includes negative region.
We are interested in  a negative potential term $V(\phi)=-\alpha^2\phi^6<0$ violating the WEC in the Einstein-minimally coupled scalar theory without  introducing any coupling function $f(\phi)$.
Hence, the tachyonic instability is absent.  In this case,  we  may obtain a single branch of scalarized black holes in Einstein-minimally coupled scalar theory.
Carrying out  stability  analysis for  scalarized black holes by adopting radial perturbations, we find that  all  scalarized black holes belonging to a single branch are unstable.
This implies that even though one obtains easily scalarized black hole from the simplest scalar-tensor action, they are hard  to  survive for further implications.

The present work is  motivated partly from the hairy black hole solutions in Einstein-Weyl-massive conformally coupled scalar theory where the tachyonic scalar mass has played an important role in obtaining scalar hairy black holes~\cite{Sultana:2020pcc}.  In this case, in the absence of tachyonic mass term, we have obtained the non-BBMB black hole solution
in the new massive conformal gravity~\cite{Myung:2019adj}.  Also, this work is motivated partly from finding  scalarized black hole  in the Einstein-minimally coupled scalar theory with  an asymmetric scalar  potential  which contains a negative region, indicating a violation of the WEC~\cite{Corichi:2005pa,Gubser:2005ih,Chew:2022enh,Chew:2024rin}. Besides, the asymmetric potential has also been employed to construct the fermionic stars \cite{DelGrosso:2023dmv,Berti:2024moe}. A symmetric potential violating the WEC was also introduced to obtain the scalarized black holes~\cite{Chew:2023olq}, which can be smoothly connected with the counterpart gravitating scalaron in the small horizon limit \cite{Chew:2024bec}. Note that this gravitating scalaron can possess the positive Arnowitt-Deser-Misner (ADM) mass but previously another type of gravitating scalaron with negative ADM mass was constructed by employing the Higgs-like potential with a phantom field \cite{Dzhunushaliev:2008bq}, hence this demonstrates that the use of phantom field can be avoided for the construction of gravitating scalaron.  The no-hair theorem suggests that asymptotically flat black holes with scalar hair  do not exist if a scalar matter  satisfies the WEC~\cite{Bekenstein:1995un}.
Hence, if the WEC for a scalar matter is violated, scalarized black holes could be found from the Einstein-minimally coupled scalar theory.
As far as we know, our model  corresponds to a simplest model which violates the WEC  because of a negative potential and thus, it could induce scalarized black holes without introducing a non-minimal scalar coupling term. Other types of scalarized black holes in the Einstein-minimally coupled scalar theory can be found in Ref. \cite{Bechmann:1995sa,Dennhardt:1996cz,Bronnikov:2001ah,Martinez:2004nb,Nikonov:2008zz,Anabalon:2012ih,Stashko:2017rkg,Gao:2021ubl,Karakasis:2023ljt,Atmaja:2023ing,Li:2023tkw,Rao:2024fox}. Besides, recently this similar concept has been adopted to construct a traversable wormhole in the Einstein-3-form theory with the Higgs-like potential, which is sufficient to violate the null energy condition where the phantom field is no longer needed and the kinetic term still can remain the correct sign \cite{Bouhmadi-Lopez:2021zwt}.

\section{No tachyonic instability of Schwarzschild black hole}

We start with  the  Einstein-minimally coupled scalar theory defined by
\begin{eqnarray}S_{\rm Es}=\int d^4 x\sqrt{-g}
\Big[\frac{R}{16\pi G}-\frac{1}{2}\nabla_\mu\phi \nabla^\mu\phi-V(\phi)\Big],
\label{Ecs}
\end{eqnarray}
where $V(\phi)$ is a negative scalar potential
\begin{equation}
V(\phi)=- \alpha^2 \phi^6 \label{s-potent}
\end{equation}
with $\alpha>0$.

The Einstein equation is derived from (\ref{Ecs}) as
\begin{eqnarray} \label{nequa1}
R_{\mu\nu}- \frac{1}{2}g_{\mu\nu}R&=&8\pi G\Big[\nabla_\mu\phi \nabla_\nu\phi-g_{\mu\nu}\Big(\frac{1}{2}\nabla_\alpha\phi \nabla^\alpha\phi+V(\phi)\Big)\Big].
\end{eqnarray}
On the other hand, the scalar
equation is given by
\begin{equation} \label{scalar-eq}
\nabla^2\phi=\frac{dV(\phi)}{d\phi}.
\end{equation}
Considering
\begin{equation}
\bar{R}_{\mu\rho\nu\sigma}\not=0,~~\bar{R}_{\mu\nu}=0,~~\bar{R}=0,~~\bar{\phi}=0,
\end{equation}
Eqs. (\ref{nequa1}) and (\ref{scalar-eq}) imply  the
GR (Schwarzschild) black hole solution
\begin{equation} \label{schw}
ds^2_{\rm Sch}=\bar{g}_{\mu\nu}dx^\mu
dx^\nu=-f(r)dt^2+\frac{dr^2}{f(r)}+r^2d\Omega^2_2
\end{equation}
with the metric function
\begin{equation} \label{num}
f(r)=1-\frac{r_0}{r}.
\end{equation}
Here the event horizon appears at $r_H=r_0$. We note that $\bar{\phi}=$ const is  not allowed for getting a GR black hole.

Now, let us   introduce the  scalar and metric perturbations around the
Schwarzschild  black hole
\begin{eqnarray} \label{m-p}
g_{\mu\nu}=\bar{g}_{\mu\nu}+h_{\mu\nu},~~\phi=0+\delta \phi.
\end{eqnarray}
Firstly, we consider the scalar perturbation.
Considering (\ref{scalar-eq}),  its linearized scalar equation is
 given by
\begin{equation}\label{nlsca}
\bar{\nabla}^2\delta \phi=0.
\end{equation}
Reminding the spherically symmetric
background (\ref{schw}), it is convenient to separate the scalar perturbation
into modes
\begin{equation}
\delta \phi(t,r,\theta,\varphi)=e^{-i\omega t } Y_{l m
}(\theta,\varphi)\frac{u(r)}{r}\,, \label{sep}
\end{equation}
where $Y_{l m}(\theta, \varphi)$ is spherical  harmonics with $-m\le l
\le m$. Introducing a tortoise coordinate $r_*=r+r_0 \ln(r/r_0-1)$ defined by $dr_*(r)=dr/f(r)$,  a radial part of the wave
equation leads to the Schr\"{o}dinger-type equation as
\begin{equation}
\frac{d^2u}{dr_*^2} +[\omega^2-V_{\rm s}(r)] u(r)=0,
\end{equation}
where the scalar potential $V_{\rm s}(r)$  takes the form
\begin{equation}
V_{\rm s}(r)=f(r)\Big[\frac{r_0}{r^3}+\frac{l(l+1)}{r^2}\Big],
\end{equation}
which is always positive definite for any $l$ outside the horizon. Therefore, there is no tachyonic instability and the Schwarzschild black hole is stable against scalar perturbation.

Now, we briefly mention  the tensor perturbation.
The linearized Einstein equation around the Schwarzschild black hole
is simply given by
\begin{eqnarray} \label{nlin-eq}
\delta G_{\mu\nu}=0
\end{eqnarray}
where the linearized Einstein tensor is expressed in terms of linearized Ricci tensor and Ricci scalar as
\begin{eqnarray}
\delta G_{\mu\nu}&=&\delta R_{\mu\nu}-\frac{1}{2} \delta
R\bar{g}_{\mu\nu},
\label{ein-t} \\
\delta
R_{\mu\nu}&=&\frac{1}{2}\Big(\bar{\nabla}^{\rho}\bar{\nabla}_{\mu}h_{\nu\rho}+
\bar{\nabla}^{\rho}\bar{\nabla}_{\nu}h_{\mu\rho}-\bar{\nabla}^2h_{\mu\nu}-\bar{\nabla}_{\mu}
\bar{\nabla}_{\nu}h\Big), \label{ricc-t} \\
\delta R&=& \bar{g}^{\mu\nu}\delta R_{\mu\nu}= \bar{\nabla}^\mu
\bar{\nabla}^\nu h_{\mu\nu}-\bar{\nabla}^2 h \label{Ricc-s}
\end{eqnarray}
with $h=h^\rho~_\rho$.
Taking the trace of the linearized
Einstein equation (\ref{nlin-eq}), one has
\begin{equation}
\delta R=0.
\end{equation}
Plugging
$\delta R=0$  into Eq. (\ref{nlin-eq}) leads to the
linearized GR equation for the linearized Ricci tensor
\begin{equation} \label{slin-eq}
\delta R_{\mu\nu}=0.
\end{equation}
It is well known that the Schwarzschild black hole is stable against  metric perturbation $h_{\mu\nu}$ in the GR ~\cite{Regge:1957td,Zerilli:1970se,Vishveshwara:1970cc,Kwon:1986dw}.
This implies that the linearized stability analysis has no prediction on exploring  scalarized black holes.

\section{Scalarized black holes}
Even though the Schwarzschild black hole appears stable against scalar perturbation, we may obtain scalarized black holes because the potential in (\ref{s-potent}) violates the WEC.
First of all, we introduce a spherically symmetric metric  to construct the scalarized  black holes solutions
\begin{equation}  \label{line-element}
ds^2_{\rm Sb} = - N(r) e^{-2 \sigma(r)} dt^2 + \frac{dr^2}{N(r)} + r^2 d\Omega^2_2
\end{equation}
with $N(r)=1-2 m(r)/r$ where $m(r)$ denotes the Misner-Sharp mass function. We note  that $m(\infty)=M$, the total mass of the configuration.
%Hereafter,  we choose $\alpha=1$ for simplicity.

Plugging  Eq.(\ref{line-element}) into  (\ref{nequa1}) and (\ref{scalar-eq}),  three differential equations for  metric functions ($m,\sigma$) and scalar ($\phi$)  are found as
\begin{equation} \label{three-eq}
m' = 2 \pi G r^2 \left( N \phi'^2 + 2 V \right), \quad \sigma' = - 4 \pi G r \phi'^2,   \quad
\left(  e^{- \sigma} r^2 N \phi' \right)' = e^{- \sigma} r^2  \frac{d V}{d \phi},
\end{equation}
where the prime ($'$) denotes the derivative with respect to the radial coordinate $r$.

 At this stage, one needs to know the asymptotic behaviour of these functions at the horizon and the infinity to construct globally defined  black hole solutions.
 Near the horizon ($r\simeq r_H$), the leading forms  in the series expansion are given by
\begin{align}
 m(r) &= \frac{r_H}{2}+ m_1 (r-r_H) + O\left( (r-r_H)^2 \right), \\
\sigma(r) &= \sigma_H + \sigma_1   (r-r_H) + O\left( (r-r_H)^2 \right), \\
 \phi(r) &= \phi_H +  \phi_{H,1}  (r-r_H) + O\left( (r-r_H)^2 \right),
\end{align}
where the coefficients are given by
\begin{equation}
   m_1 = 4 \pi G r^2_H  V(\phi_H), \quad  \sigma_1 = -  4 \pi G r_H \phi^2_{H,1}, \quad   \phi_{H,1}= \frac{r_H V'(\phi_H)}{1-8 \pi G r_H^2 V(\phi_H)}.
\end{equation}
Here $\sigma_H$ and $\phi_H$ are the values of $\sigma$ and $\phi$ at the horizon. We note that   the denominator of $\phi_{H,1}$ should satisfy the condition of  $1-8 \pi G r_H^2 V(\phi_H)\not=0$  to keep $\sigma(r)$ and $\phi(r)$ should be finite at the horizon.

Asymptotic expansions of the metric and scalar functions at infinity take  the forms  when   imposing  asymptotical flatness and a vanishing scalar as
\begin{align}
    m(r) &= M  -  \frac{ D^2 }{4 r} -  \frac{ M D^2 }{4 r^2} - \frac{D^2}{3 r^3} \left( 2 D^4 + M^2 \right) + \mathcal{O}(r^{-4}) \,, \\
    \sigma(r) &=  \frac{ D^2  }{4 r^2}  +  \frac{2 M D^2}{3 r^3} + \left( \frac{3D^6}{2} - \frac{D^4}{8} + 3 M^2 D^2 \right) \frac{1}{2 r^4} + \mathcal{O}(r^{-5}) \,,  \\
    \phi(r) &= -\frac{ D  }{ r} - \frac{DM}{r^2} + \left( -D^4 + \frac{D^2}{12} - \frac{4M^2}{3} \right) \frac{D}{r^3} + \mathcal{O}(r^{-4}) \,,
\end{align}
where  $M$ and $D$ represent the ADM mass and scalar charge of the hairy black hole, respectively.

Finally, introducing two dimensionless parameters
\begin{equation}
 r \rightarrow  \frac{r}{\alpha \sqrt{8\pi G} },\quad m \rightarrow  \frac{m}{\alpha \sqrt{8\pi G}}\,,
\end{equation}
three equations in Eq.(\ref{three-eq}) can be  solved by an ODE solver package Colsys  when adapting  the Newton-Raphson method to solve the boundary value problem for three coupled  nonlinear differential equations~\cite{Ascher:1979iha}.
We employ a compactified coordinate $x=1-r_H/r (x\in [0,1])$ for constructing hairy black holes in the numerics.  Here, we are left with the three parameters $(\phi_H,\sigma_H, r_H)$ where $\sigma_H$  is  determined when imposing $\sigma(\infty)=0$ while $r_H$ could be determined when all solutions satisfy the boundary conditions.  In this case, the state of a scalarized black hole depends only on $\phi_H$ which means that different scalarized black holes are encoded in different $\phi_H$.
When the scalar field is zero on the horizon, the corresponding configuration is solely given by  the Schwarzschild black hole.
 However, when the scalar field on the horizon $\phi_H$ is non-zero and then, one  increases it, a branch of  hairy black holes bifurcates and behaves quite differently from the Schwarzschild black hole.
 To represent this behavior, we introduce reduced area $a_H$ and reduced temperature $t_H$  by making use of  area of horizon $A_H$ and Hawking temperature $T_H$ of hairy black holes as
 \begin{equation}
a_H=\frac{A_H}{16\pi M^2}, \quad t_H= 8\pi T_M M \quad {\rm with} \quad   A_H=4\pi r_H^2,\quad T_H=\frac{1}{4\pi}N'(r_H)e^{-\sigma_H}.
\end{equation}
\begin{figure}
\centering
\mbox{
(a)
\includegraphics[angle =-90,scale=0.29]{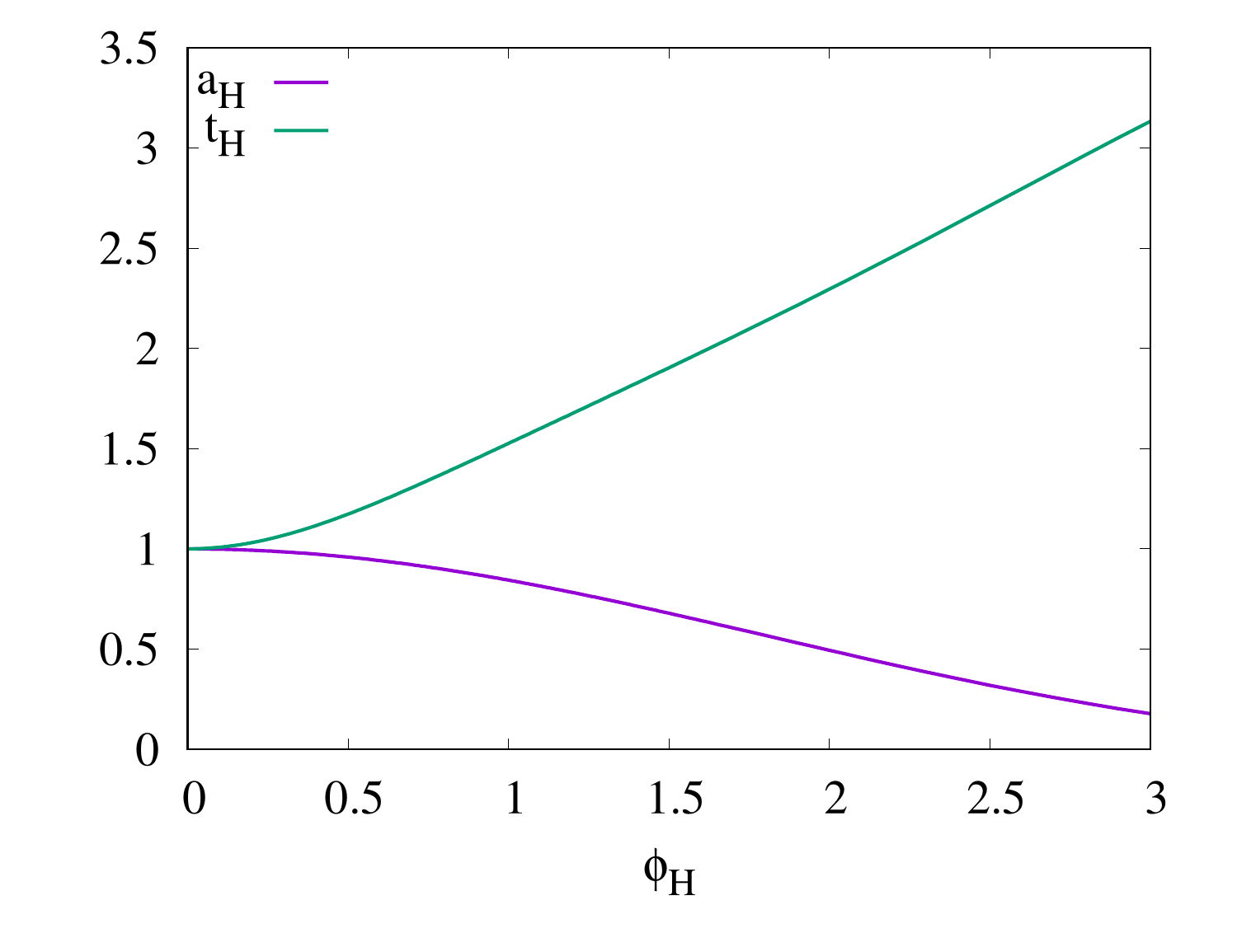}
(b)
\includegraphics[angle =-90,scale=0.29]{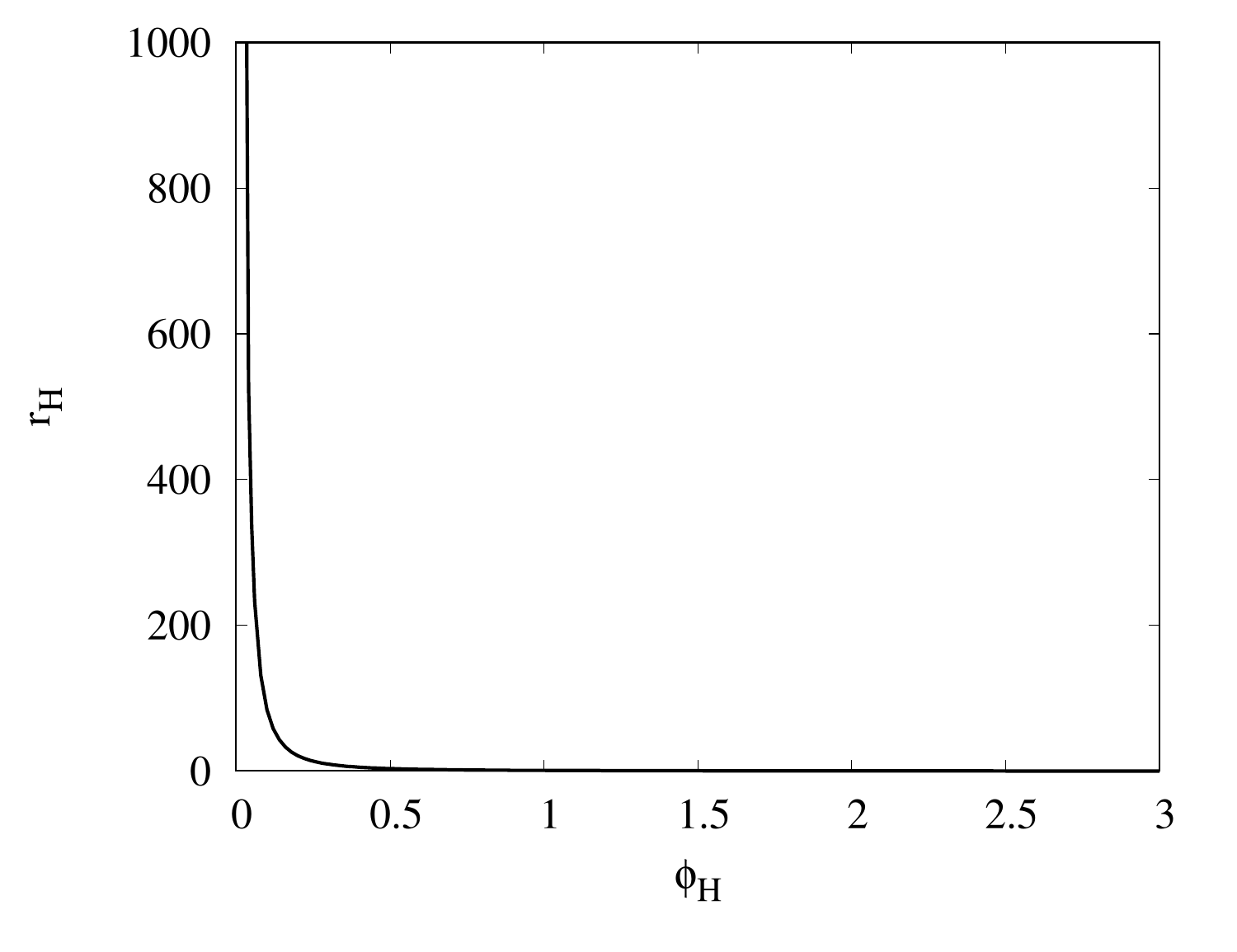}
}
\caption{Two basic quantities of hairy black hole: (a)  reduced area of horizon $a_H$  and  reduced Hawking temperature $t_H$ as functions of the scalar field at the horizon $\phi_H$.
Two are starting with $a_H=t_H=1$ for the Schwarzschild black hole at $\phi_H=0$. (b) horizon radius  $r_H$ as a function of  $\phi_H$.}
\label{plot_prop1}
\end{figure}
Fig. \ref{plot_prop1}(a) shows the plots of reduced area of horizon $a_H$ and reduced Hawking temperature $t_H$. We recall that two are $a_H=t_H=1$ for the Schwarzschild black hole with $\phi_H=0$.
For increasing $\phi_H$, $a_H$ decreases monotonically from unity to very close to zero whereas  $t_H$ increases monotonically from unity. Although some hairy black holes were constructed by different $V(\phi)$ \cite{Chew:2022enh}, their $a_H$ and $t_H$ behave qualitatively similar to Fig. \ref{plot_prop1}(a).
Fig. \ref{plot_prop1}(b) indicates that the radius of horizon $r_H$ of hairy black hole is inversely proportional to $\phi_H$, which means that the hairy black holes bifurcates from the Schwarzschild black hole with a very large value of $r_H$ and finally, $r_H$ could shrink to zero as $\phi_H$ increases.  Here, both $r_H$ and $\phi_H$ could take any arbitrary positive real values.

\begin{figure}
\centering
\mbox{
(a)
\includegraphics[angle =-90,scale=0.29]{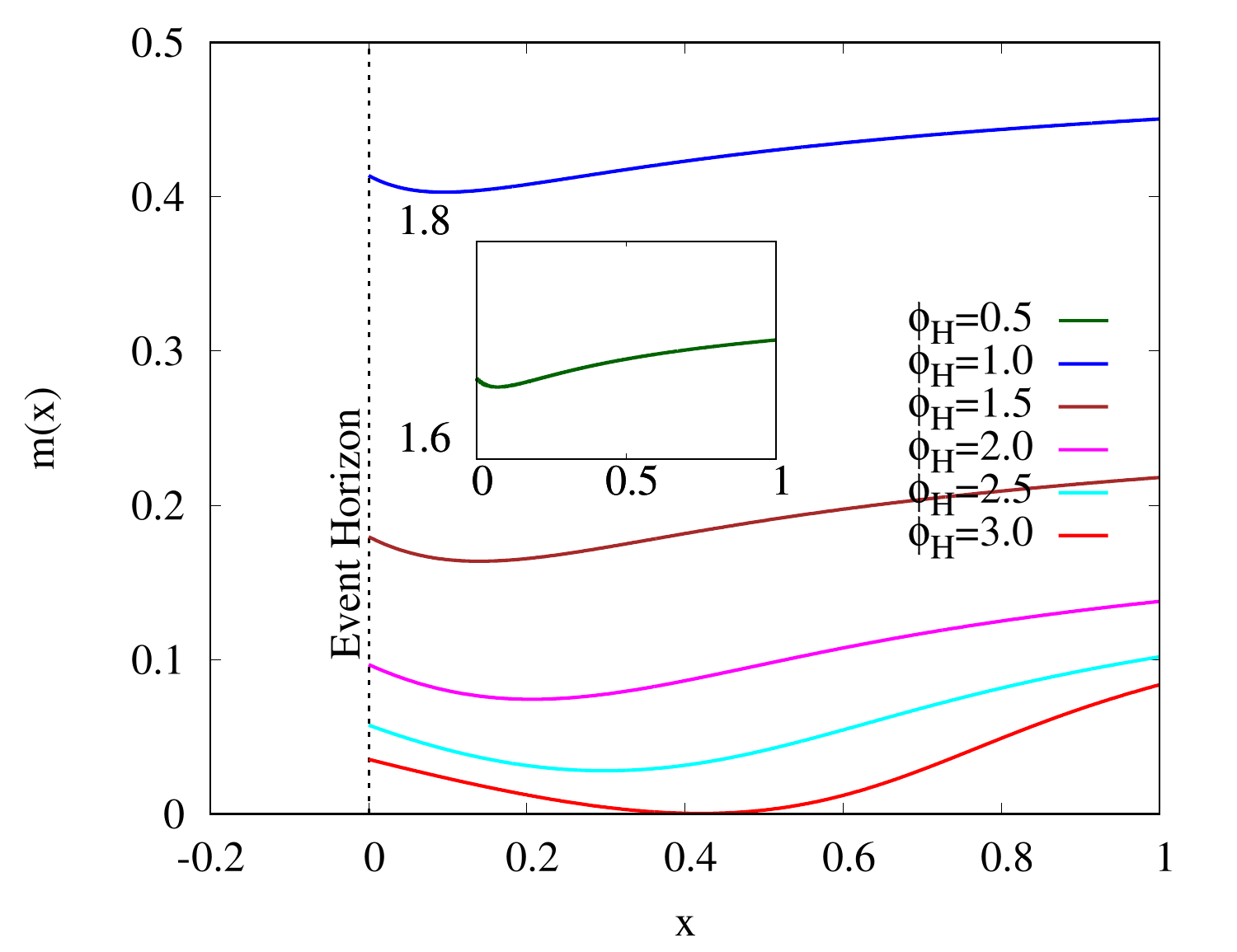}
(b)
\includegraphics[angle =-90,scale=0.29]{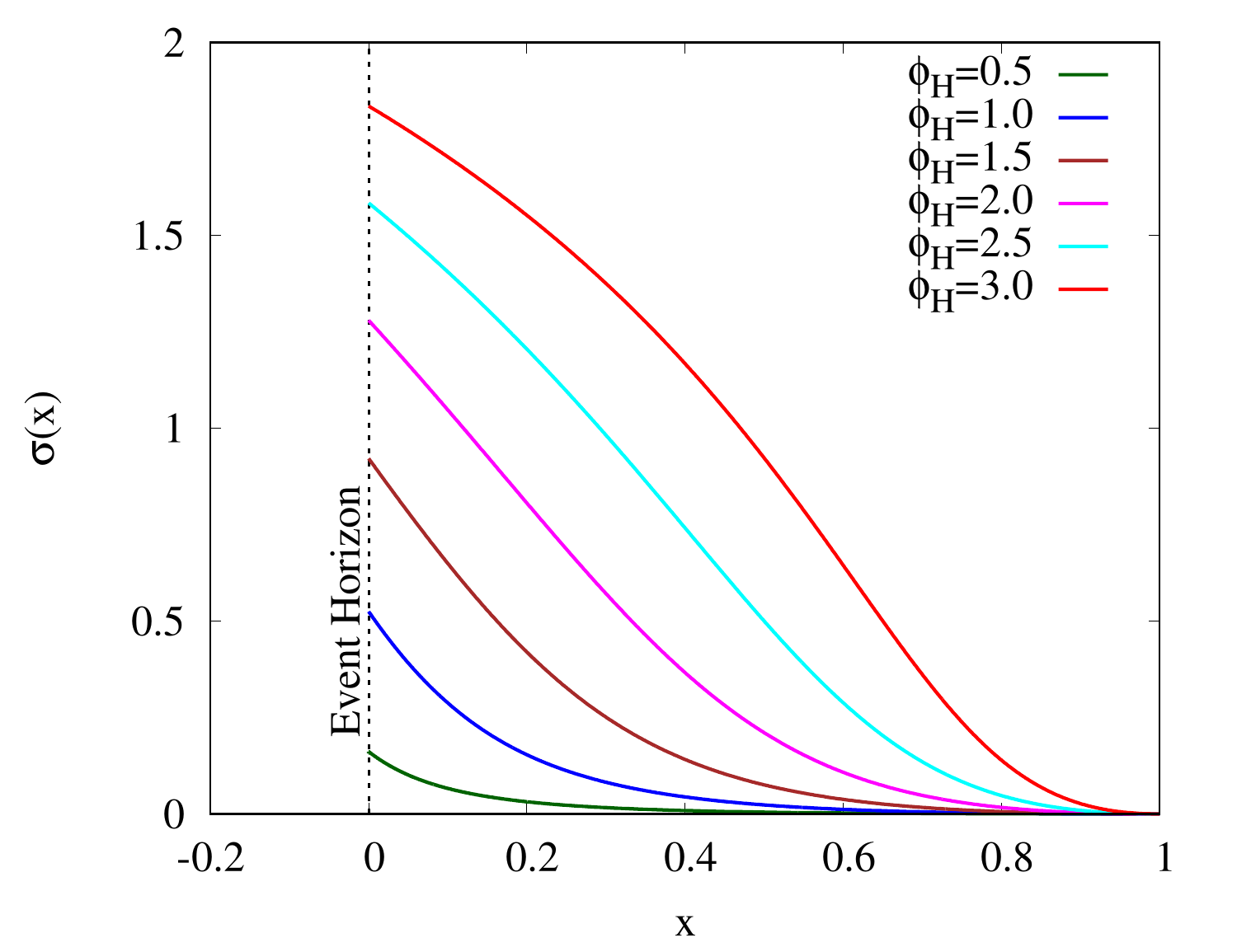}
}
 \mbox{
 (c)
 \includegraphics[angle =-90,scale=0.29]{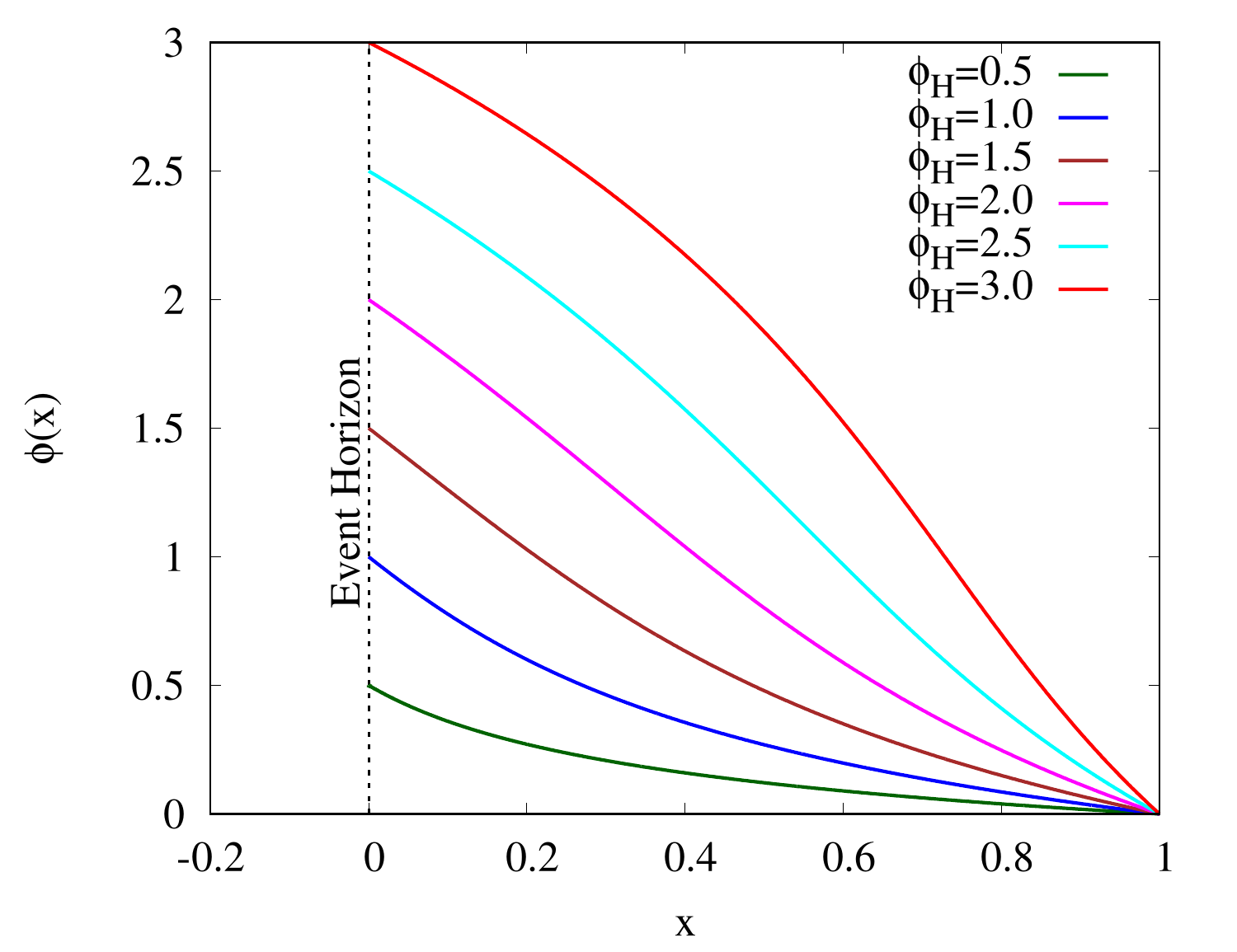}
 }
\caption{Six solutions with different $\phi_H$ of hairy black hole in the compactified coordinate $x$: (a) mass function $m(x)$ contains an inset for $\phi_H=0.5$ (b) metric function $\sigma(x)$, and (c) scalar hair $\phi(x)$.}
\label{plot_sol}
\end{figure}
Now, we are in a position to present the numerical solutions and discuss their properties.
Fig. \ref{plot_sol} exhibits six solutions of hairy black hole  with a choice of six different $\phi_H$ as functions of the compactified coordinate  $x$, where they are regular everywhere outside and on the horizon.
 The functions $\sigma(x)$ and $\phi(x)$ decrease monotonically from its maximum value at the horizon to zero at the infinity.  However, the mass function $m(x)$ possesses a local minimum which moves away from the horizon to the infinity as $\phi_H$ increases.

We might  understand the presence of scalarized black holes by observing the WEC.
\begin{figure}
\centering
\mbox{
%(a)
\includegraphics[angle =-90,scale=0.28]{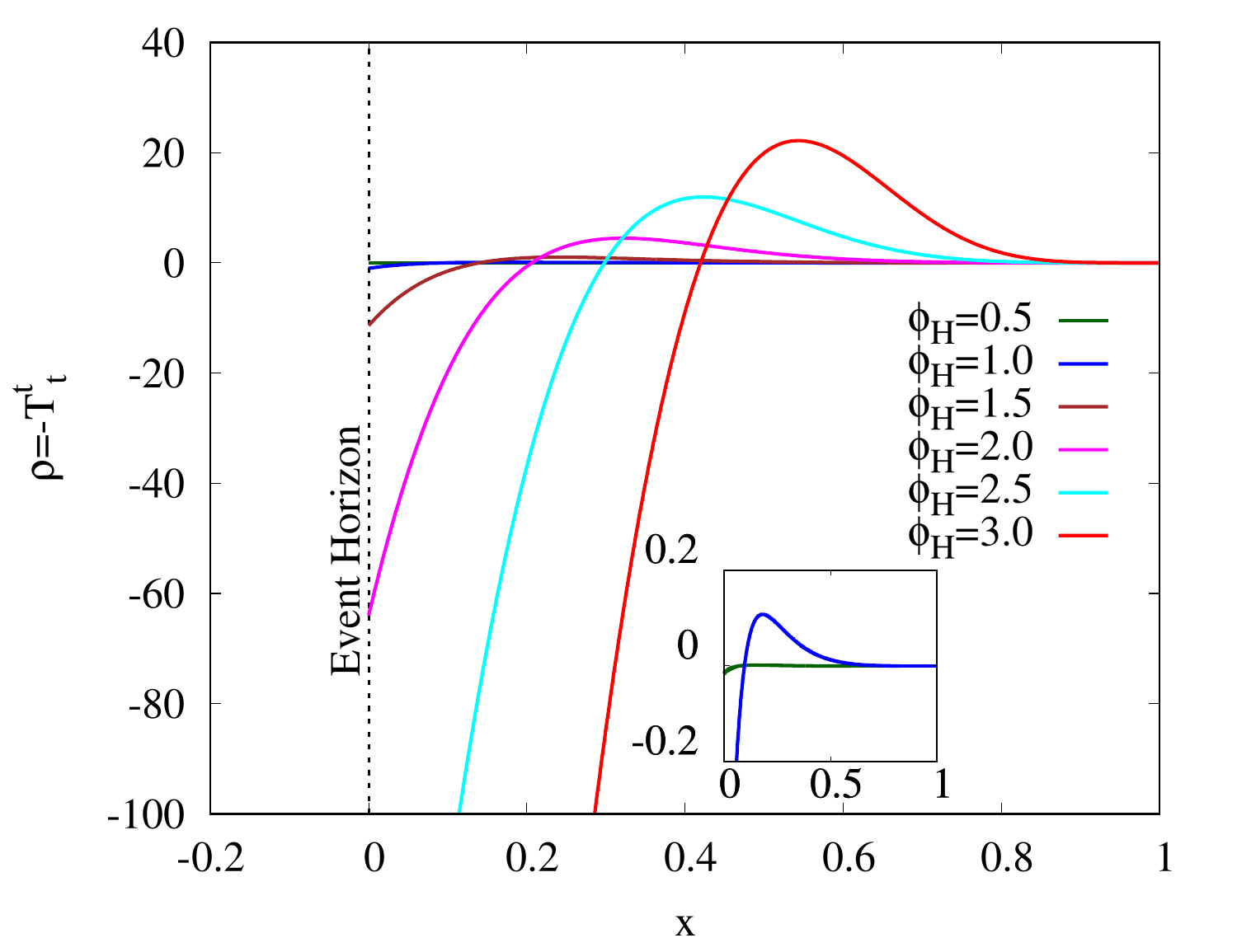}
%(b)
%\includegraphics[angle =-90,scale=0.26]{plot_p6BH_rho}
}
% \mbox{
% (c)
% \includegraphics[angle =-90,scale=0.26]{plot_symm_scalaron_Vphoton}
%(b)
% \includegraphics[angle =-90,scale=0.26]{plot_symm_scalaron_VISCO2}
% }
%
\caption{The energy density $\rho$ for the hairy black holes with six $\phi_H$ in the compactified coordinate $x\in[0,1]$. The inset shows $\rho$ for $\phi_H=0.5, 1.0$.}
\label{plot_ec}
\end{figure}
We examine the energy condition of the hairy black hole as shown in Fig.~\ref{plot_ec}.  The WEC  is described by the energy density $\rho=-T^t\,_t=N\phi'^2/2+V(\phi)$  which is  negative near the horizon. This shows violation of the WEC clearly. One may evade  no-hair theorem  if the scalar matter  does not satisfy the WEC~\cite{Bekenstein:1995un}. This implies the presence of scalarized black holes.  $\rho$ at the horizon decreases very sharply with the increase of $\phi_H$.  Besides, $\rho$ possesses a local maximum which is located exactly at the local minimum of $m(x)$, and it moves further away from the horizon to the infinity as $\phi_H$ increases.

\section{Radial perturbations around scalarized black holes}
For further implications of scalarized black holes, we need to perform  stability test for them.
For this purpose, we introduce radial perturbations defined by
\begin{eqnarray}
 ds^2 &=& - N(r) e^{-2 \sigma(r)} \left[  1 + \epsilon e^{-i \omega t} F_t(r)  \right] dt^2 + \frac{1}{N(r)} \left[ 1+ \epsilon e^{-i \omega t} F_r(r)   \right] dr^2 + r^2 d\Omega_2^2, \label{per-eq1} \\
 \Phi &=& \phi(r) + \epsilon  \Phi_1 (r) e^{-i \omega t},\label{per-eq2}
\end{eqnarray}
where $ F_t(r)$,  $F_r(r)$, and $\Phi_1 (r)$ are three perturbed fields.

Substituting Eqs.(\ref{per-eq1}) and (\ref{per-eq2}) into   (\ref{nequa1}) and (\ref{scalar-eq}), we obtain three linearized equations
\begin{eqnarray}
 F_r &=& 8 \pi G r\phi'  \Phi_1, \label{ODEper1} \\
 F'_t &=& - F'_r + 16 \pi G r \phi'  \Phi'_1, \label{ODEper2} \\
\Phi''_1 &=& \left(  \sigma' - \frac{N'}{N} -\frac{2}{r} \right)  \Phi'_1 + \left(  \frac{1}{N}  V''(\phi)  - \omega^2 \frac{e^{2 \sigma}}{N^2}  \right)  \Phi_1 +V'(\phi)\frac{F_r}{N} + \phi'\frac{ F'_r - F'_t }{2}. \label{ODEper3}
\end{eqnarray}
We eliminate the last two terms in Eq.(\ref{ODEper3})  by making  use of  the first two equations to obtain an independent scalar equation.
Then, we can transform $\Phi''_1 $ to  Schr\"odinger-like equation by introducing $Z(r) = r \Phi_1(r)$  and  a tortoise coordinate $r_*$ defined  by  $r_* =\int dr [\frac{e^{\sigma}}{N}]$ as
\begin{equation}
 \frac{d^2 Z}{d r_{*}^2} + \Big[ \omega^2 - V_R(r) \Big] Z =0
 \label{Z_radial}
\end{equation}
with the effective potential
\begin{equation}
    V_{\rm R}(r) = N e^{-2 \sigma} \left[ \frac{N}{r} \left( \frac{N'}{N} - \sigma'  \right)  - 8 \pi G r N \phi'^2 \left(  \frac{N'}{N} + \frac{1}{r} - \sigma' \right) + 16 \pi G r \phi' V'(\phi) + V''(\phi) \right].
 \label{V_R}
\end{equation}
\begin{figure}
\centering
\mbox{
(a)
\includegraphics[angle =-90,scale=0.28]{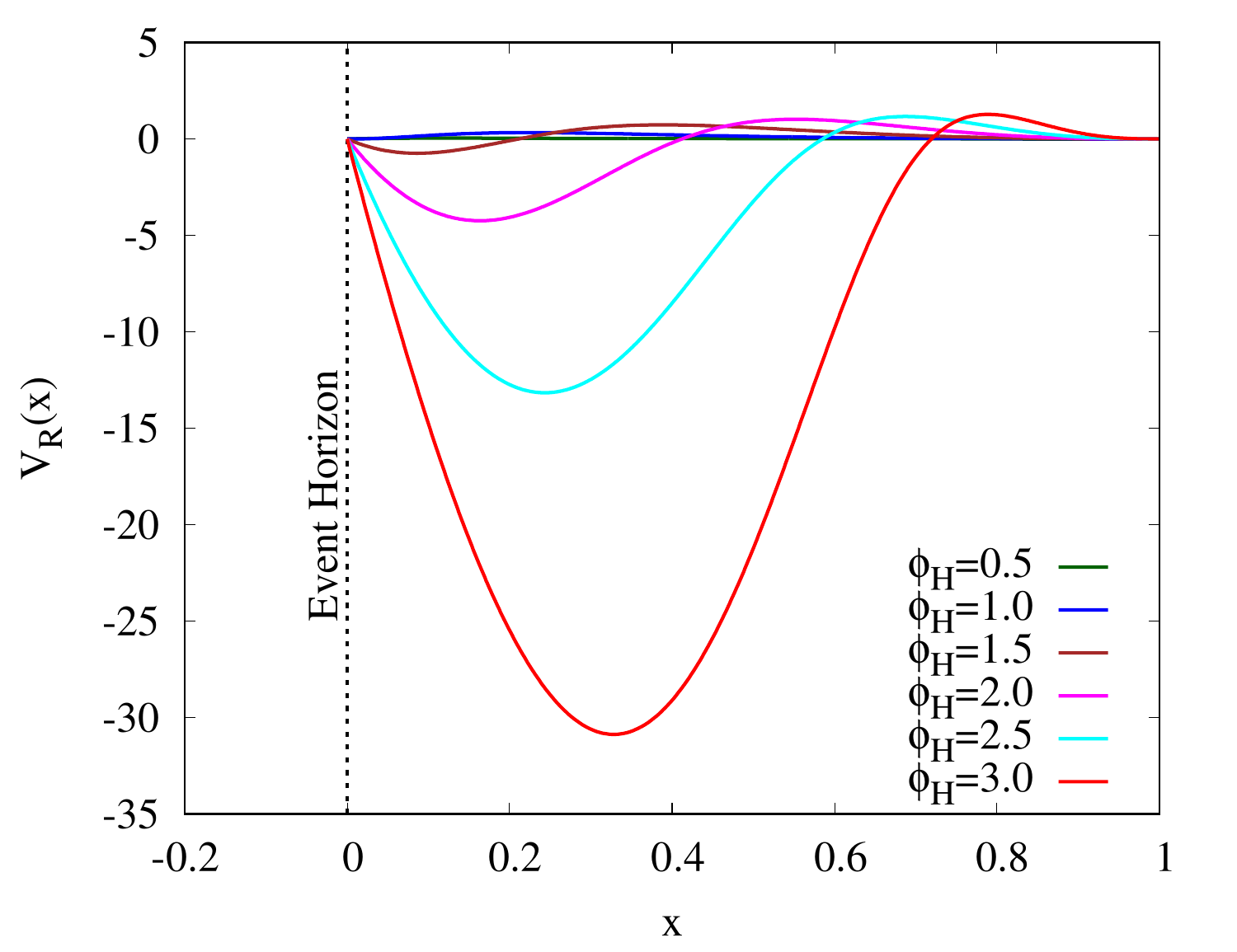}
(b)
\includegraphics[angle =-90,scale=0.28]{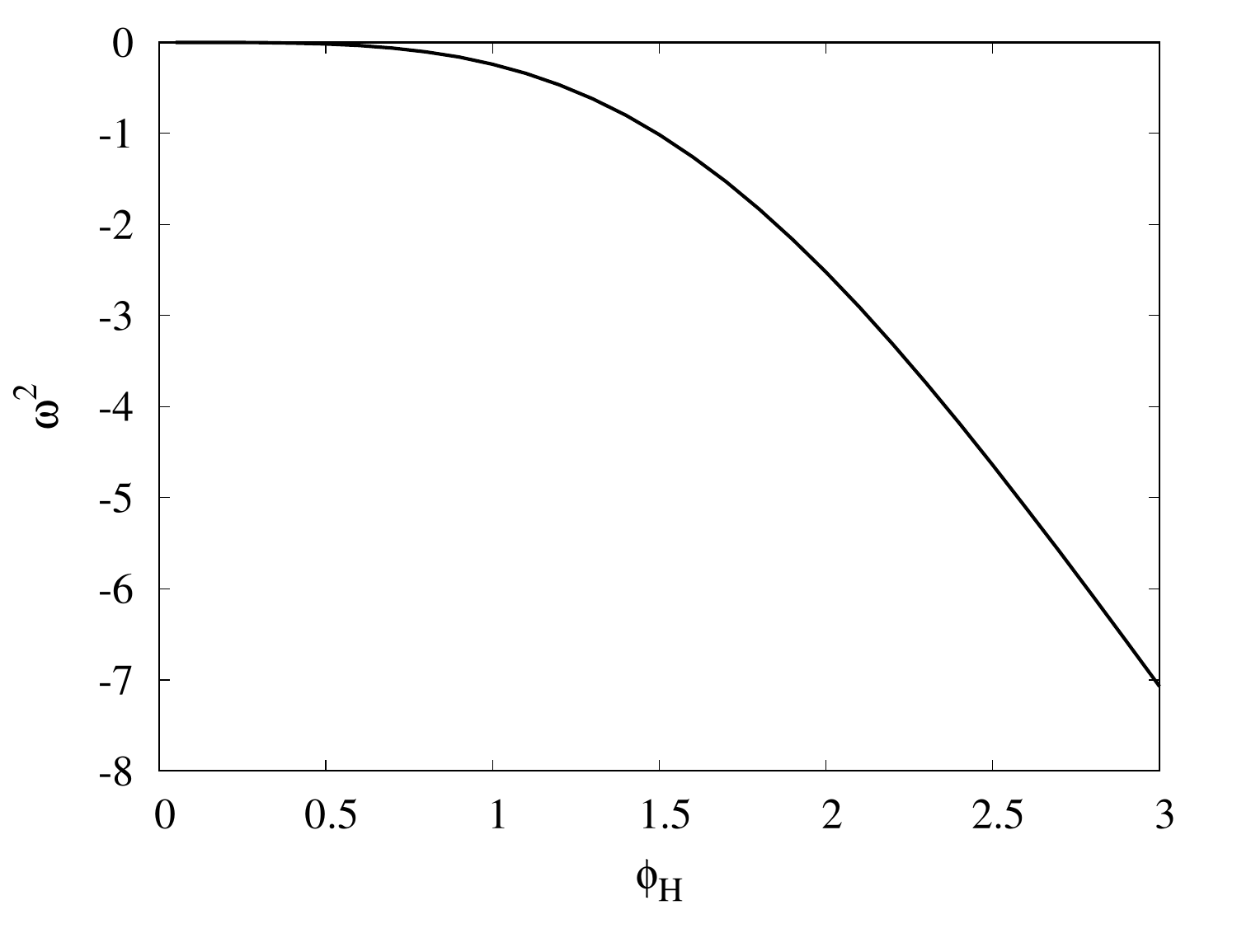}
}
\caption{(a) The effective potential $V_R(x)$ with six different $\phi_H$ in the compactified coordinate $x\in[0,1]$ for the Schr\"odinger-like equation. (b) The eigenvalue $\omega^2$ for unstable modes as a function of  $\phi_H$. It is obvious that $\omega^2<0$.  }
\label{plot_modes}
\end{figure}

We wish to  perform the linear stability of the hairy black hole. Before we procced, it is worth mentioning that  as Fig.~\ref{plot_modes}(a) is shown,  the effective potentials  $V_R(x)$ with six different $\phi_H$  are always  negative in some regions of $x$, implying  the possible presence of  unstable modes.
We note  that solving  Eq.(\ref{Z_radial}) corresponds  to handling  an eigenvalue problem. Hence, we obtain the radial mode numerically by using COLSYS to solve it  with $\omega^2$ as an eigenvalue.
For black holes, we impose that the perturbation fields vanish at two boundaries, $Z(r_H)=Z(\infty)=0$.  In the numerics, we introduce an auxiliary equation of $\frac{d}{d r} [\omega^2]=0$. This allows us to impose an additional condition of $Z(r_p)=1$ at some point $r_p$, which is typically located at the middle of the horizon and infinity.   This allows us to obtain a nontrivial and normalizable solution for $Z$, since Eq.~\eqref{Z_radial} is homogeneous.  The eigenvalue $\omega^2$ is determined  automatically when $Z$ satisfies all  asymptotic boundary conditions.
Accordingly, Fig.~\ref{plot_modes}(b) indicates that the  $\omega^2 <0$ (unstable modes)  decreases with the increase of $\phi_H$  where the scalar perturbation increases exponentially with time.
The perturbation $Z$ is unstable because  $\omega^2=-\Omega^2<0$ where the time-dependent perturbation ($e^{\Omega t}$) grows exponentially with time.
This implies that all scalarized black holes belonging to a single branch are unstable.

\section{Discussions}
It is clear that the tachyonic instability as the onset of spontaneous scalarization indicates infinite branches of scalarized black holes.
We have explored  a single branch of scalarized black holes in the Einstein-minimally coupled scalar theory with a negative potential $V(\phi)=-\alpha^2\phi^6$.
Here, the tachyonic instability is absent  and thus, it plays no role in predicting infinite branches of scalarized black holes. In this case, one could not meet  a condition for  spontaneous scalarization, but one meets a condition for nonlinear scalarization to obtain a single branch of scalarized black holes in the Einstein-Gauss-Bonnet-scalar theory with a coupling function $f(\phi)$.
This suggests that tachyonic instability is a necessary condition to obtain infinite branches of scalarized black holes.

To generate  a single branch of scalarized black hole, there are some sources of nonlinear instability~\cite{Blazquez-Salcedo:2022omw,Doneva:2021tvn,Pombo:2023lxg,Zhang:2023jei}, conformal scalar coupling~\cite{Bocharova:1970skc,Bekenstein:1974sf,Zou:2020zxq}, superradiant instability~\cite{Herdeiro:2014goa}, and violation of the WEC~\cite{Corichi:2005pa,Gubser:2005ih,Chew:2022enh,Chew:2024rin}.
In our work, it is important to note that the negative  scalar potential with $\alpha=1$ violates the weak energy condition (WEC).
It is well-known that if the WEC for a scalar matter is violated,  scalarized black holes could be found from the Einstein-minimally coupled scalar theory without introducing any scalar coupling function $f(\phi)$ to matter.
Thus, we have obtained the black hole solutions with scalar hair  by solving three nonlinear  equations.
It includes  a single branch of scalarized black holes only because  tachyonic intability is absent.  Different scalarized black holes are encoded in different $\phi_H$ because we have chosen $\alpha=1$.
Furthermore, we have studied their thermodynamic properties by introducing reduced horizon area and Hawking temperature.

Then, we have performed  the stability  analysis  for scalarized black holes by adopting radial perturbations.
It turned out that  six  scalarized black holes with six different $\phi_H$  belonging to a single branch are unstable.
Therefore, it is unlikely that this scalarized black hole can be considered as the astrophysical black holes such as M87 and SgrA*, since the detection on their existence from the astrophysical signatures could be very challenging.

Finally, it would be interesting to obtain the other solutions of scalarized black holes by choosing a more simpler form of the potential, for instance $V(\phi)=-m^2_\phi \phi^2,-\Lambda \phi^4$ with $m^2_\phi$ and $\Lambda$ are positive constants, since the previous analysis of Eqs.(8) and (9) in~\cite{Chew:2023olq}  doesn't rule out  the possible existence of scalarized black hole solutions. Therefore, it is meaningful to investigate such possibilities and report them in the future.

 \vspace{2cm}

{\bf Acknowledgments} \\
 \vspace{1cm}

XYC acknowledges the support from the starting grant of Jiangsu University of Science and Technology (JUST).

\end{document}